%% file: PRL_main.tex
\documentclass[%
reprint,
 amsmath,amssymb,
prl,
]{revtex4-2}

\usepackage{graphicx}
\usepackage{dcolumn}
\usepackage{bm}
\usepackage{hyperref}
\usepackage{enumitem}
\usepackage{xcolor}

\def\Xmax{$X_\text{max}$\phantom{ }}
\def\XmaxFULLSTOP{$X_\text{max}$}

\begin{document}

\preprint{APS/123-QED}

\title{Demonstrating Agreement between Radio and Fluorescence Measurements of the Depth of Maximum of Extensive Air Showers at the Pierre Auger Observatory}

\author{\input{latex_auger_authorlist_authors}}
\affiliation{}
\collaboration{Pierre Auger Collaboration}

\author{\phantom{1}}
\affiliation{\input{latex_auger_authorlist_institutions.tex}}%

\date{\today}

\begin{abstract}
\newpage
We show, for the first time, radio measurements of the depth of shower maximum ($X_\text{max}$) of air showers induced by cosmic rays that are compared to measurements of the established fluorescence method at the same location. Using measurements at the Pierre Auger Observatory we show full compatibility between our radio and the previously published fluorescence data set, and between a subset of air showers observed simultaneously with both radio and fluorescence techniques, a measurement setup unique to the Pierre Auger Observatory. Furthermore, we show radio \Xmax resolution as a function of energy and demonstrate the ability to make competitive high-resolution \Xmax measurements with even a sparse radio array. With this, we show that the radio technique is capable of cosmic-ray mass composition studies, both at Auger and at other experiments. 
\end{abstract}

\maketitle
The origin and nature of cosmic rays has been one of the driving questions in astroparticle physics in the past century. Especially for \textit{ultra-high energy cosmic rays} much remains to be discovered about their sources, their acceleration mechanisms, and how they propagate. A particularly important range of cosmic-ray energies to investigate is the so-called \textit{transition region}. There the sources of cosmic rays are expected to transition from Galactic to extragalactic origin, which is commonly expected to occur in the energy range between $10^{17}$ and $10^{19}$\,eV~\cite{ref:UHECRWhitePaper}. Current efforts in this regime focus on measuring the cosmic-ray flux, the arrival direction, and the composition of cosmic-ray primaries. Of these, mass composition is particularly important to distinguish between different possible source models.

The Pierre Auger Observatory~\cite{ref:augernim} in Argentina, covering $3000$\,km$^2$, is the largest facility dedicated to detecting ultra-high-energy cosmic rays (UHECRs). The primary components are an array of $1660$ water-Cherenkov detectors, also called the \textit{surface detector} (SD) and $27$ fluorescence telescopes, known as the \textit{fluorescence detector} (FD) that overlook the SD. The observatory also has an array of radio detectors, the \textit{Auger Engineering Radio Array} (AERA)~\cite{ref:AERA_Main}, located within the grid of the SD and close to one of the FD sites. AERA was constructed to measure the radio signals produced in extensive air showers at energies between $10^{17}$ and $10^{19}$\,eV. It thus probes the transition region with independent and complementary measurements to those made with fluorescence light, air-Cherenkov light, and secondary particles of air showers. The technique of radio detection of cosmic rays has made great steps in the past twenty years providing understanding of the emission mechanisms, the implementation in simulation codes, and the reconstruction of shower properties~\cite{ref:histlopes,ref:codalema,ref:MGMR,ref:zhaires,ref:MicroCoREAS,ref:2dldf,ref:EnergyScalePRL} (see also ~\cite{ref:HuegeReview,ref:Schroder,ref:Astro2020} for extensive reviews). 

Radio emission in air showers is produced by time-varying currents from the movement of electrons and positrons. These arise from acceleration in the magnetic field of the Earth and ionization of the atmosphere while the shower develops. The currents give rise to electromagnetic radiation at frequencies, predominantly, in the MHz to GHz regime that arrives on the ground as a short pulse of a few nanoseconds. The frequency spectrum and spatial distribution are governed by the fact that the source moves relativistically in a medium with a refractive index gradient, which leads to a Cherenkov-like time compression. By sampling the radio-emission footprint over an extended area with an array of radio antennas, one can reconstruct the properties of the air shower and derive information about the primary cosmic ray. For example, the arrival direction of the cosmic ray can be reconstructed based on the arrival times of the signals in multiple antennas and the strength of the radio emission footprint scales with the energy of the air shower~\cite{ref:Astro2020,ref:EnergyScalePRL,ref:EnergyScalePRD}. The general shape of the footprint also changes with the particle type of the primary cosmic ray. This is because a heavier primary particle (e.g., an iron nucleus) essentially behaves like a superposition of lower-energy protons that interact earlier in the atmosphere than a single proton with all the energy. The heavier particle will thus produce a wider radio emission footprint on the ground. Therefore, the shape of the footprint is a mass-sensitive probe for the primary particle type. We don't directly observe the particle type, but it is strongly related to the atmospheric depth $X$ where the shower is maximally developed, the \textit{depth of the shower maximum $X_\text{max}$}, which we can observe. Hence, \Xmax is used as the main probe in this work to investigate the types of particles inducing the air-shower signals measured by AERA. 

In this work we present the results of a technique to measure \XmaxFULLSTOP, developed for AERA, using data measured over $7$~years. We compare this to measurements from the FD to show compatibility and, in addition, perform a direct comparison of \Xmax of showers measured simultaneously by both detectors. Next, we evaluate the resolution of the method to demonstrate the competitiveness of the radio method. Finally, we compare these results to other experiments and discuss the implications.

\begin{figure*}[!ht]
\includegraphics[width=2\columnwidth]{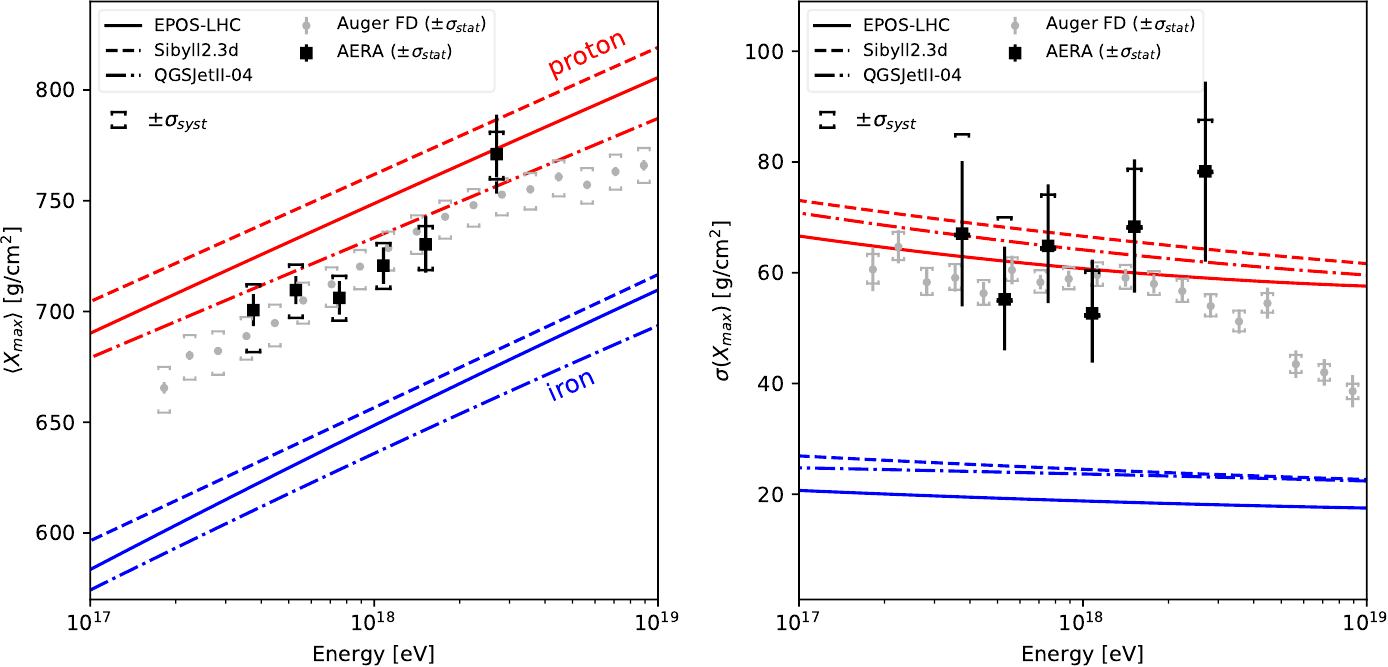}
\caption{\label{fig:Elongation}Mean (left) and standard deviation (right) of the $X_\textup{max}$ distribution as measured by AERA in this work (black). The results are compared to predictions from {\sc CORSIKA} air-shower simulations for three hadronic interaction models (lines) for proton (red) and iron (blue) mass compositions~\cite{ref:hadr_epos,ref:hadr_sibyll,ref:hadr_qgs,ref:POA_Composition_ICRC2019} and compared to measurements by the Auger FD~\cite{ref:POA_Composition_ICRC2019}. The statistical uncertainties on the mean and width of the measurements are plotted as error bars and the systematic uncertainties with capped markers.}
\end{figure*}

\textit{The \Xmax Distribution.}---In Fig.\,\ref{fig:Elongation} we show the first two central moments of the distributions of reconstructed \Xmax values (as a function of the SD energy~\cite{ref:SD_energyscale}) resulting from $594$ measured air showers. For this, we have used the state-of-the-art air-shower simulation code ({\sc CORSIKA} v7.7100~\cite{ref:Corsikamain} with radio extension {\sc CoREAS}~\cite{ref:MicroCoREAS}) to generate an ensemble of $27$ simulated air showers for each of our measured air showers. To achieve the highest precision possible we use a model of the atmosphere~\cite{ref:gdastool,ref:gdasinoffline} and geomagnetic field~\cite{ref:magneticmodel} at the time and location of each shower. These simulations are generated such that they cover the \Xmax phase space. We then compare the measured radio signals to the simulated signals to derive the \Xmax value that best represents the measurements. Details on the reconstruction method, which builds upon~\cite{ref:LOFARresults1,ref:LOFARresults0}, are presented in an accompanying publication~\cite{ref:AERAXmaxPRD}. The $594$ showers have been selected to have energies above $E=10^{17.5}$\,eV, the threshold for full efficiency of the SD particle trigger~\cite{ref:Infill_Efficiency_v2,ref:Infill_Efficiency_v3}, and to be detectable by AERA for any realistically occurring \Xmax value (i.e., an acceptance cut for radio)~\cite{ref:AERAXmaxPRD}. With the results, we demonstrate that the AERA measurements of the first and second moment of the \Xmax distribution (black markers) are compatible with the measurements of the fluorescence telescopes at the Pierre Auger Observatory (gray markers). Note that while for $\langle X_\text{max}\rangle$ a mixed composition will result in values in between the lines for a pure proton and a pure iron composition, a mixed composition can result in $\sigma(X_\text{max})$ values even larger than those of a pure proton composition. The statistical agreement of the results of AERA and the FD provides independent support for the validity of the FD measurements~\cite{ref:POA_Composition_ICRC2019} and shows that the radio method is able to perform the same measurements. It also confirms the validity of the microscopic radio-emission simulations of {\sc CoREAS}. The comparison of radio and fluorescence might also provide a way in the future to improve constraints on the systematic uncertainties of the fluorescence method. For example, by lowering the uncertainties on atmospheric corrections. 

\begin{figure}[!ht]
\includegraphics[width=\columnwidth]{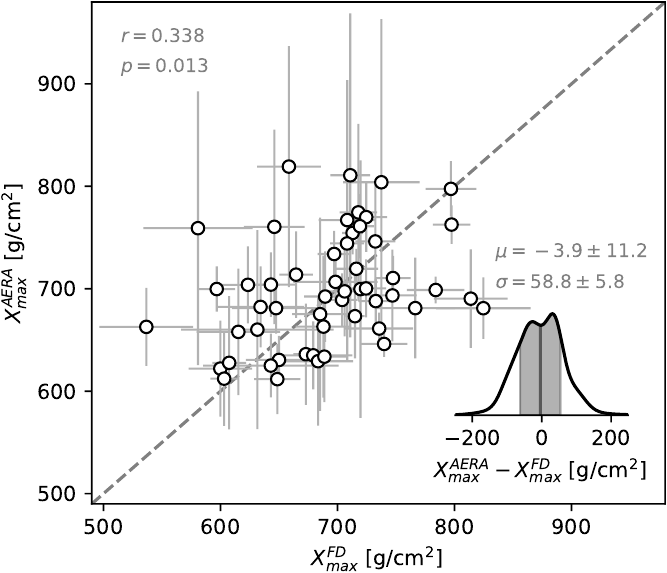}
\caption{\label{fig:FDresultsScatter}Comparison of \Xmax for showers measured simultaneously by both AERA and the FD. A diagonal line is shown to guide the eye. Shown at the top is the Pearson correlation coefficient $r$ with corresponding $p$ value (the probability to obtain an $r$ of at least that value from uncorrelated data). Shown at the bottom is the distribution (kernel density estimation) of the differences with mean $\mu$ and spread $\sigma$.}
\end{figure}

\textit{Direct Comparison with Hybrid Radio-Fluorescence Measurements.}--- We can also make a direct comparison between the two \Xmax reconstruction techniques at Auger, using a subset of $53$ air showers (predominantly between $10^{17.5}$ and $10^{18}$\,eV), that were measured simultaneously by both the FD and AERA. When comparing the \Xmax values on an event-by-event basis (Fig.\,\ref{fig:FDresultsScatter}) we find an average difference of $\langle X_\textup{max}^\textup{AERA}-X_\textup{max}^\textup{FD} \rangle=-3.9\pm11.2$\,g\,cm$^{-2}$, demonstrating there to be no significant bias. The distribution of the differences is compatible with a Gaussian distribution with the combined \Xmax resolution of our method and the FD ($53.3\pm5.7$\,g\,cm$^{-2}$ versus the distribution width of $58.8\pm5.8$\,g\,cm$^{-2}$). Additionally, the average difference shows no significant change when applying, for example, cuts on energy or \Xmax resolution, indicating this set of hybrid showers is well-behaved. The average difference further strengthens the agreement between the fluorescence and radio methods as it shows agreement not just on the mean \Xmax versus energy between two data sets, but also on an event-to-event level where the effects of event-selection bias are absent. 

Furthermore, the agreement between the two methods directly illustrates that both the FD and radio \Xmax reconstructions are well-understood. The fluorescence method involves imaging the trajectory of the air shower. When one accounts for the attenuation of the light one can extract the depth in the atmosphere where the fluorescence emission is strongest, corresponding to \XmaxFULLSTOP. The radio technique in contrast is not affected by attenuation, yet other effects play a role. The coherence of the radio signal is a key factor as it strongly affects what we observe in our antennas. Thus, the spatial distribution of particles in the shower down to the scales set by our highest frequency ($80$\,MHz, corresponding to $3.75$\,m) is directly probed. Furthermore, the radio emission is the result of two emission mechanisms that interfere with each other (arising from time-varying transverse and longitudinal currents) and it is in addition affected by the refractive index gradient of the atmosphere. Because of this complexity, we have used air-shower simulations to obtain \Xmax by comparing the measured and simulated radio signals in our antennas. So, when we are comparing the \Xmax measurements of the two techniques, we not only test that all of these effects are accounted for correctly, but we inherently also test the implementation of the radio-emission calculation in simulations (both the simulation of the electromagnetic cascade as well as the radio emission in a discretized classical electrodynamics calculation). The agreement on \Xmax by AERA and the FD thus strongly suggests that all of these aspects are well under control.

\begin{figure}[!t]
\includegraphics[width=\columnwidth]{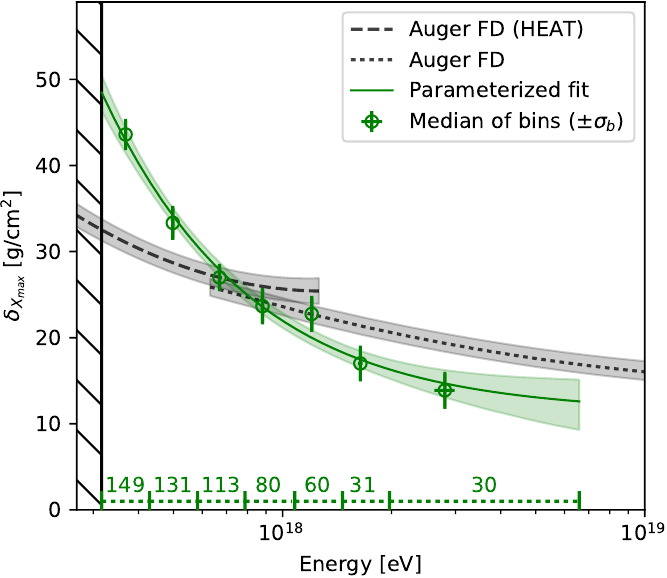}
\caption{\label{fig:RDresolution}Resolution of the $X_\textup{max}$ reconstruction method, $\delta_{X_\text{max}}$, as a function of energy in units of column density. The median values of the uncertainties on \Xmax (circles with uncertainties $\sigma_b$ from bootstrap resampling) for our set of showers are shown per energy bin along with the parametrized fit [Eq.~(\ref{eq:EnergyResolution})] of the resolution of \Xmax (solid line with $1\sigma$-confidence bands). Also shown are the resolutions achieved by the Auger fluorescence telescopes~\cite{ref:FDXmaxSyst}. The black hatched region at low energy indicates the cut on energy for this AERA analysis. The size of the energy bins with the number of showers per bin is inset at the bottom of the figure.}
\end{figure}

\textit{The \Xmax Resolution.}---We determined an uncertainty for each reconstructed \Xmax value based on the reconstruction of simulated showers, allowing us to directly evaluate the resolution of our method. In Fig.\,\ref{fig:RDresolution} we show the median \Xmax resolution versus cosmic-ray energy $E$ (green points), demonstrating that we are able to reach a resolution of better than $15$\,g\,cm$^{-2}$ at the highest energies ($13.9\pm2.0$\,g\,cm$^{-2}$ for the last bin). Towards lower energies, the resolution becomes worse, mainly because of the weaker radio signals at lower energies (leading to lower signal-to-noise ratios in our antennas). For comparison we also show the resolution obtained by the fluorescence telescopes at the Pierre Auger Observatory~\cite{ref:FDXmaxSyst}, demonstrating the competitiveness of the radio technique over a wide energy range. Because of the large set of showers, we are also able to evaluate the energy dependence of the \Xmax resolution. We parameterize our resolution $\delta_{X_\text{max}}$ as a function of energy (green line) inspired by the energy resolution of electromagnetic calorimeters ~\cite{ref:CalorimetryHandbook} and similar to the shapes used for the FD:
    \begin{equation}
    \delta_{X_\text{max}} = a \cdot \sqrt{\frac{10^{18} \text{eV}}{E}} \oplus b \cdot \frac{10^{18}\text{eV}}{E} \oplus c,  \label{eq:EnergyResolution}
    \end{equation}
where $a=14.0\pm6.8$\,g\,cm$^{-2}$, $b=12.7\pm2.5$\,g\,cm$^{-2}$, and $c=11.2\pm4.7$\,g\,cm$^{-2}$ are free parameters, and $\oplus$ indicates the quadratic sum. The $c$ parameter provides a prediction of the potential resolution that our method might be able to reach for AERA data. For radio experiments with a denser antenna spacing or experiments with lower ambient noise conditions one might reasonably expect this resolution to be even better. For example, LOFAR reported an average resolution of $19$\,g\,cm$^{-2}$ using a similar method~\cite{ref:LOFAR2021} and simulation studies for the upcoming Square Kilometer Array suggest an average resolution of $6-8$\,g\,cm$^{-2}$ could be reached~\cite{ref:SKAxmaxres}. In all likelihood their respective resolutions will improve with energy similar to the trend shown for AERA, making the radio technique very competitive for precision \Xmax measurements.

\begin{figure}[!ht]
\includegraphics[width=\columnwidth]{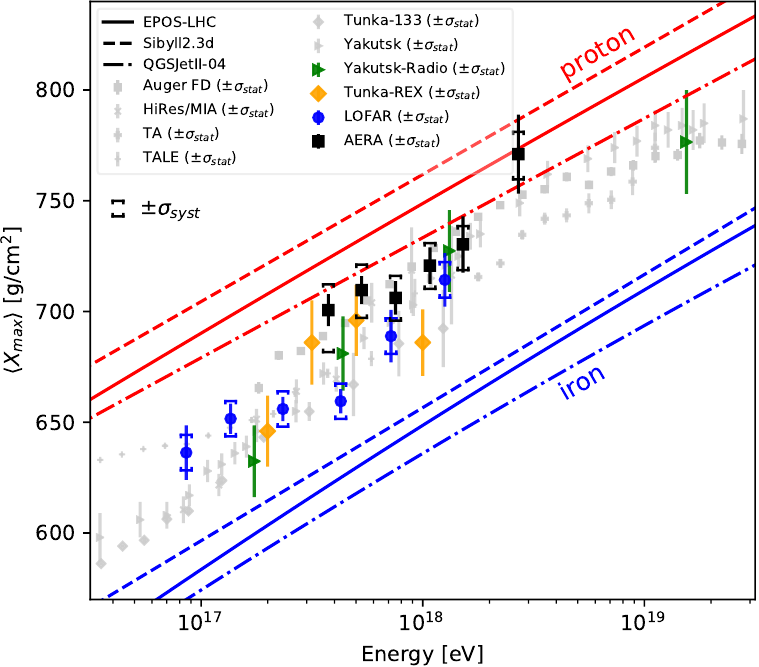}
\caption{\label{fig:ElongationRadio}Mean of the $X_\textup{max}$ distribution as measured by AERA in this work (black). The results are compared to predictions from air-shower simulations for multiple hadronic interaction models (lines) for proton (red) and iron (blue) mass compositions~\cite{ref:hadr_epos,ref:hadr_sibyll,ref:hadr_qgs,ref:POA_Composition_ICRC2019} and compared to measurements by LOFAR~\cite{ref:LOFAR2021}, Tunka-Rex~\cite{ref:Xmax_TunkaREX}, Yakutsk-Radio~\cite{ref:YakutskXmax2019}, and Auger FD~\cite{ref:POA_Composition_ICRC2019}. Note that the Yakutsk-Radio results do not account for aperture effects on the same level as the other experiments. Colors have been used to highlight the measurements with the radio technique. The statistical uncertainties on the measurements are shown as vertical bars and for radio the systematic uncertainties, if available, are shown with caps.}
\end{figure}

\textit{Comparison to other Experiments.}---In Fig.\,\ref{fig:ElongationRadio} we show our $\langle X_\text{max}\rangle$ results together with various results from previous works. Measurements by other experiments that use the radio technique to measure $X_\text{max}$ are highlighted in color. 
In the past, Tunka-Rex~\cite{ref:Xmax_TunkaREX}, Yakutsk-Radio~\cite{ref:YakutskXmax2019}, and LOFAR~\cite{ref:LOFAR2021} (and its prototype LOPES~\cite{ref:LOPESXmax}) have shown \Xmax measurements, but it has been challenging to make significant statements on the compatibility of the radio technique with fluorescence and air-Cherenkov light measurements. This is because these experiments either didn't have a second technique to directly compare to or due to a combination of large statistical uncertainties and limited investigation of detector-specific systematic uncertainties. It is difficult to make statements on the compatibility of AERA and Tunka-Rex or Yakutsk-Radio without a full picture of those systematic uncertainties, but there do not seem to be significant discrepancies within their statistical uncertainties  (note that the highest energy bin of Tunka-Rex only contains $10$ showers, hence its deviation with AERA is arguably not significant). However, the LOFAR measurements include a detailed estimation of systematic uncertainties, have much smaller statistical uncertainties than Tunka-Rex or Yakutsk-Radio, and share many similarities with AERA in the method to reconstruct \XmaxFULLSTOP, so we can compare these results to the FD and AERA results.

We note that the difference between the Auger FD and LOFAR measurements, as can be seen in Fig.\,\ref{fig:ElongationRadio}, previously left open the possibility of a systematic shift in \Xmax due to an inherent difference between radio and fluorescence techniques. However, the AERA \Xmax results now show no significant bias w.r.t.\ the fluorescence results, not when comparing their full data sets nor on an event-to-event level. Additionally, a study of the compatibility of the full shape of the \Xmax distribution as measured by AERA and the Auger FD, available in~\cite{ref:AERAXmaxPRD}, also finds no significant discrepancies within uncertainties. This strongly suggests that the differences between Auger and LOFAR must be either physical (e.g., due to differences in the magnetic field or atmospheric conditions, their altitudes, or their southern versus northern exposure) or systematic (e.g., due to the event selection or reconstruction), but not inherent to either the radio or fluorescence techniques.

At higher energies, a seemingly similar difference in $\langle X_\text{max}\rangle$ (both in magnitude and direction) can be observed between the fluorescence results of Auger (gray squares) and TA (gray plus markers). However, a detailed comparison by an Auger-TA working group has found that, given the known selection bias in the TA data, this difference is compatible within uncertainties~\cite{ref:AugerTAWGXmax2023}. This comparison only covers energies above $10^{18.2}$\,eV, so does not overlap with the LOFAR data. An AERA-LOFAR working group has started looking into their apparent differences, investigating, for example, differences in event selection, \Xmax reconstruction method, and energy scale. Regardless, a deeper comparison of AERA and LOFAR data opens a new way to try to understand and reduce systematic uncertainties on air-shower and cosmic-ray parameters. Furthermore, the combination of fluorescence and radio measurements, linked by hybrid detectors such as at Auger, might resolve or constrain differences even more. 

\textit{Conclusions}.---In this work, we have used $7$~years of AERA data to investigate the depth of maximum of extensive air showers at energies where the cosmic-ray origin is expected to transition from Galactic to extragalactic sources. We show our \Xmax results to be in agreement with the results of the fluorescence telescopes at the Pierre Auger Observatory. In addition, this compatibility is also demonstrated on an event-by-event level with simultaneous radio and fluorescence measurements of the same air showers. With our method, we are able to achieve competitive high-resolution \Xmax reconstructions, reaching resolutions near $15$\,g\,cm$^{-2}$ at the highest energies. With this, we have demonstrated that the reconstruction of \Xmax at AERA is both well-understood and competitive with established methods and ready to be used in future experiments.
\input{latex_auger_acknowledgments}


\end{document}

%% file: latex_auger_authorlist_authors.tex
A.~Abdul Halim$^{13}$,
P.~Abreu$^{73}$,
M.~Aglietta$^{55,53}$,
I.~Allekotte$^{1}$,
K.~Almeida Cheminant$^{71}$,
A.~Almela$^{7,12}$,
R.~Aloisio$^{46,47}$,
J.~Alvarez-Mu\~niz$^{79}$,
J.~Ammerman Yebra$^{79}$,
G.A.~Anastasi$^{55,53}$
% ...

%% file: latex_auger_authorlist_institutions.tex
\begin{description}[labelsep=0.2em,align=right,labelwidth=0.7em,labelindent=0em,leftmargin=2em,noitemsep]
\item[$^{1}$] Centro At\'omico Bariloche and Instituto Balseiro (CNEA-UNCuyo-CONICET), San Carlos de Bariloche, Argentina
\item[$^{2}$] Departamento de F\'\i{}sica and Departamento de Ciencias de la Atm\'osfera y los Oc\'eanos, FCEyN, Universidad de Buenos Aires and CONICET, Buenos Aires, Argentina
\item[$^{3}$] IFLP, Universidad Nacional de La Plata and CONICET, La Plata, Argentina
\item[$^{4}$] Instituto de Astronom\'\i{}a y F\'\i{}sica del Espacio (IAFE, CONICET-UBA), Buenos Aires, Argentina
\item[$^{5}$] Instituto de F\'\i{}sica de Rosario (IFIR) -- CONICET/U.N.R.\ and Facultad de Ciencias Bioqu\'\i{}micas y Farmac\'euticas U.N.R., Rosario, Argentina
\item[$^{6}$] Instituto de Tecnolog\'\i{}as en Detecci\'on y Astropart\'\i{}culas (CNEA, CONICET, UNSAM), and Universidad Tecnol\'ogica Nacional -- Facultad Regional Mendoza (CONICET/CNEA), Mendoza, Argentina
\item[$^{7}$] Instituto de Tecnolog\'\i{}as en Detecci\'on y Astropart\'\i{}culas (CNEA, CONICET, UNSAM), Buenos Aires, Argentina
\item[$^{8}$] International Center of Advanced Studies and Instituto de Ciencias F\'\i{}sicas, ECyT-UNSAM and CONICET, Campus Miguelete -- San Mart\'\i{}n, Buenos Aires, Argentina
\item[$^{9}$] Laboratorio Atm\'osfera -- Departamento de Investigaciones en L\'aseres y sus Aplicaciones -- UNIDEF (CITEDEF-CONICET), Argentina
\item[$^{10}$] Observatorio Pierre Auger, Malarg\"ue, Argentina
\item[$^{11}$] Observatorio Pierre Auger and Comisi\'on Nacional de Energ\'\i{}a At\'omica, Malarg\"ue, Argentina
\item[$^{12}$] Universidad Tecnol\'ogica Nacional -- Facultad Regional Buenos Aires, Buenos Aires, Argentina
\item[$^{13}$] University of Adelaide, Adelaide, S.A., Australia
\item[$^{14}$] Universit\'e Libre de Bruxelles (ULB), Brussels, Belgium
\item[$^{15}$] Vrije Universiteit Brussels, Brussels, Belgium
\item[$^{16}$] Centro Brasileiro de Pesquisas Fisicas, Rio de Janeiro, RJ, Brazil
\item[$^{17}$] Centro Federal de Educa\c{c}\~ao Tecnol\'ogica Celso Suckow da Fonseca, Petropolis, Brazil
\item[$^{18}$] Instituto Federal de Educa\c{c}\~ao, Ci\^encia e Tecnologia do Rio de Janeiro (IFRJ), Brazil
\item[$^{19}$] Universidade de S\~ao Paulo, Escola de Engenharia de Lorena, Lorena, SP, Brazil
\item[$^{20}$] Universidade de S\~ao Paulo, Instituto de F\'\i{}sica de S\~ao Carlos, S\~ao Carlos, SP, Brazil
\item[$^{21}$] Universidade de S\~ao Paulo, Instituto de F\'\i{}sica, S\~ao Paulo, SP, Brazil
\item[$^{22}$] Universidade Estadual de Campinas, IFGW, Campinas, SP, Brazil
\item[$^{23}$] Universidade Estadual de Feira de Santana, Feira de Santana, Brazil
\item[$^{24}$] Universidade Federal de Campina Grande, Centro de Ciencias e Tecnologia, Campina Grande, Brazil
\item[$^{25}$] Universidade Federal do ABC, Santo Andr\'e, SP, Brazil
\item[$^{26}$] Universidade Federal do Paran\'a, Setor Palotina, Palotina, Brazil
\item[$^{27}$] Universidade Federal do Rio de Janeiro, Instituto de F\'\i{}sica, Rio de Janeiro, RJ, Brazil
\item[$^{28}$] Universidade Federal do Rio de Janeiro (UFRJ), Observat\'orio do Valongo, Rio de Janeiro, RJ, Brazil
\item[$^{29}$] Universidade Federal Fluminense, EEIMVR, Volta Redonda, RJ, Brazil
\item[$^{30}$] Universidad de Medell\'\i{}n, Medell\'\i{}n, Colombia
\item[$^{31}$] Universidad Industrial de Santander, Bucaramanga, Colombia
\item[$^{32}$] Charles University, Faculty of Mathematics and Physics, Institute of Particle and Nuclear Physics, Prague, Czech Republic
\item[$^{33}$] Institute of Physics of the Czech Academy of Sciences, Prague, Czech Republic
\item[$^{34}$] Palacky University, Olomouc, Czech Republic
\item[$^{35}$] CNRS/IN2P3, IJCLab, Universit\'e Paris-Saclay, Orsay, France
\item[$^{36}$] Laboratoire de Physique Nucl\'eaire et de Hautes Energies (LPNHE), Sorbonne Universit\'e, Universit\'e de Paris, CNRS-IN2P3, Paris, France
\item[$^{37}$] Univ.\ Grenoble Alpes, CNRS, Grenoble Institute of Engineering Univ.\ Grenoble Alpes, LPSC-IN2P3, 38000 Grenoble, France
\item[$^{38}$] Universit\'e Paris-Saclay, CNRS/IN2P3, IJCLab, Orsay, France
\item[$^{39}$] Bergische Universit\"at Wuppertal, Department of Physics, Wuppertal, Germany
\item[$^{40}$] Karlsruhe Institute of Technology (KIT), Institute for Experimental Particle Physics, Karlsruhe, Germany
\item[$^{41}$] Karlsruhe Institute of Technology (KIT), Institut f\"ur Prozessdatenverarbeitung und Elektronik, Karlsruhe, Germany
\item[$^{42}$] Karlsruhe Institute of Technology (KIT), Institute for Astroparticle Physics, Karlsruhe, Germany
\item[$^{43}$] RWTH Aachen University, III.\ Physikalisches Institut A, Aachen, Germany
\item[$^{44}$] Universit\"at Hamburg, II.\ Institut f\"ur Theoretische Physik, Hamburg, Germany
\item[$^{45}$] Universit\"at Siegen, Department Physik -- Experimentelle Teilchenphysik, Siegen, Germany
\item[$^{46}$] Gran Sasso Science Institute, L'Aquila, Italy
\item[$^{47}$] INFN Laboratori Nazionali del Gran Sasso, Assergi (L'Aquila), Italy
\item[$^{48}$] INFN, Sezione di Catania, Catania, Italy
\item[$^{49}$] INFN, Sezione di Lecce, Lecce, Italy
\item[$^{50}$] INFN, Sezione di Milano, Milano, Italy
\item[$^{51}$] INFN, Sezione di Napoli, Napoli, Italy
\item[$^{52}$] INFN, Sezione di Roma ``Tor Vergata'', Roma, Italy
\item[$^{53}$] INFN, Sezione di Torino, Torino, Italy
\item[$^{54}$] Istituto di Astrofisica Spaziale e Fisica Cosmica di Palermo (INAF), Palermo, Italy
\item[$^{55}$] Osservatorio Astrofisico di Torino (INAF), Torino, Italy
\item[$^{56}$] Politecnico di Milano, Dipartimento di Scienze e Tecnologie Aerospaziali , Milano, Italy
\item[$^{57}$] Universit\`a del Salento, Dipartimento di Matematica e Fisica ``E.\ De Giorgi'', Lecce, Italy
\item[$^{58}$] Universit\`a dell'Aquila, Dipartimento di Scienze Fisiche e Chimiche, L'Aquila, Italy
\item[$^{59}$] Universit\`a di Catania, Dipartimento di Fisica e Astronomia ``Ettore Majorana``, Catania, Italy
\item[$^{60}$] Universit\`a di Milano, Dipartimento di Fisica, Milano, Italy
\item[$^{61}$] Universit\`a di Napoli ``Federico II'', Dipartimento di Fisica ``Ettore Pancini'', Napoli, Italy
\item[$^{62}$] Universit\`a di Palermo, Dipartimento di Fisica e Chimica ''E.\ Segr\`e'', Palermo, Italy
\item[$^{63}$] Universit\`a di Roma ``Tor Vergata'', Dipartimento di Fisica, Roma, Italy
\item[$^{64}$] Universit\`a Torino, Dipartimento di Fisica, Torino, Italy
\item[$^{65}$] Benem\'erita Universidad Aut\'onoma de Puebla, Puebla, M\'exico
\item[$^{66}$] Unidad Profesional Interdisciplinaria en Ingenier\'\i{}a y Tecnolog\'\i{}as Avanzadas del Instituto Polit\'ecnico Nacional (UPIITA-IPN), M\'exico, D.F., M\'exico
\item[$^{67}$] Universidad Aut\'onoma de Chiapas, Tuxtla Guti\'errez, Chiapas, M\'exico
\item[$^{68}$] Universidad Michoacana de San Nicol\'as de Hidalgo, Morelia, Michoac\'an, M\'exico
\item[$^{69}$] Universidad Nacional Aut\'onoma de M\'exico, M\'exico, D.F., M\'exico
\item[$^{70}$] Universidad Nacional de San Agustin de Arequipa, Facultad de Ciencias Naturales y Formales, Arequipa, Peru
\item[$^{71}$] Institute of Nuclear Physics PAN, Krakow, Poland
\item[$^{72}$] University of \L{}\'od\'z, Faculty of High-Energy Astrophysics,\L{}\'od\'z, Poland
\item[$^{73}$] Laborat\'orio de Instrumenta\c{c}\~ao e F\'\i{}sica Experimental de Part\'\i{}culas -- LIP and Instituto Superior T\'ecnico -- IST, Universidade de Lisboa -- UL, Lisboa, Portugal
\item[$^{74}$] ``Horia Hulubei'' National Institute for Physics and Nuclear Engineering, Bucharest-Magurele, Romania
\item[$^{75}$] Institute of Space Science, Bucharest-Magurele, Romania
\item[$^{76}$] Center for Astrophysics and Cosmology (CAC), University of Nova Gorica, Nova Gorica, Slovenia
\item[$^{77}$] Experimental Particle Physics Department, J.\ Stefan Institute, Ljubljana, Slovenia
\item[$^{78}$] Universidad de Granada and C.A.F.P.E., Granada, Spain
\item[$^{79}$] Instituto Galego de F\'\i{}sica de Altas Enerx\'\i{}as (IGFAE), Universidade de Santiago de Compostela, Santiago de Compostela, Spain
\item[$^{80}$] IMAPP, Radboud University Nijmegen, Nijmegen, The Netherlands
\item[$^{81}$] Nationaal Instituut voor Kernfysica en Hoge Energie Fysica (NIKHEF), Science Park, Amsterdam, The Netherlands
\item[$^{82}$] Stichting Astronomisch Onderzoek in Nederland (ASTRON), Dwingeloo, The Netherlands
\item[$^{83}$] Universiteit van Amsterdam, Faculty of Science, Amsterdam, The Netherlands
\item[$^{84}$] Case Western Reserve University, Cleveland, OH, USA
\item[$^{85}$] Colorado School of Mines, Golden, CO, USA
\item[$^{86}$] Department of Physics and Astronomy, Lehman College, City University of New York, Bronx, NY, USA
\item[$^{87}$] Michigan Technological University, Houghton, MI, USA
\item[$^{88}$] New York University, New York, NY, USA
\item[$^{89}$] University of Chicago, Enrico Fermi Institute, Chicago, IL, USA
\item[$^{90}$] University of Delaware, Department of Physics and Astronomy, Bartol Research Institute, Newark, DE, USA
\item[$^{91}$] University of Wisconsin-Madison, Department of Physics and WIPAC, Madison, WI, USA
\item[] -----
\item[$^{a}$] Louisiana State University, Baton Rouge, LA, USA
\item[$^{b}$] Institut universitaire de France (IUF), France
\item[$^{c}$] also at University of Bucharest, Physics Department, Bucharest, Romania
\item[$^{d}$] School of Physics and Astronomy, University of Leeds, Leeds, United Kingdom
\item[$^{e}$] now at Agenzia Spaziale Italiana (ASI).\ Via del Politecnico 00133, Roma, Italy
\item[$^{f}$] Fermi National Accelerator Laboratory, Fermilab, Batavia, IL, USA
\item[$^{g}$] now at Graduate School of Science, Osaka Metropolitan University, Osaka, Japan
\item[$^{h}$] now at ECAP, Erlangen, Germany
\item[$^{i}$] Max-Planck-Institut f\"ur Radioastronomie, Bonn, Germany
\item[$^{j}$] also at Kapteyn Institute, University of Groningen, Groningen, The Netherlands
\item[$^{k}$] Colorado State University, Fort Collins, CO, USA
\item[$^{l}$] Pennsylvania State University, University Park, PA, USA
\end{description}

%% file: latex_auger_acknowledgments.tex
\begin{sloppypar}
\textit{Acknowledgments.}--- The successful installation, commissioning, and operation of the Pierre
Auger Observatory would not have been possible without the strong
commitment and effort from the technical and administrative staff in
Malarg\"ue. We are very grateful to the following agencies and
organizations for financial support:
\end{sloppypar}

\begin{sloppypar}
Argentina -- Comisi\'on Nacional de Energ\'\i{}a At\'omica; Agencia Nacional de
Promoci\'on Cient\'\i{}fica y Tecnol\'ogica (ANPCyT); Consejo Nacional de
Investigaciones Cient\'\i{}ficas y T\'ecnicas (CONICET); Gobierno de la
Provincia de Mendoza; Municipalidad de Malarg\"ue; NDM Holdings and Valle
Las Le\~nas; in gratitude for their continuing cooperation over land
access; Australia -- the Australian Research Council; Belgium -- Fonds
de la Recherche Scientifique (FNRS); Research Foundation Flanders (FWO),
Marie Curie Action of the European Union Grant No.~101107047; Brazil --
Conselho Nacional de Desenvolvimento Cient\'\i{}fico e Tecnol\'ogico (CNPq);
Financiadora de Estudos e Projetos (FINEP); Funda\c{c}\~ao de Amparo \`a
Pesquisa do Estado de Rio de Janeiro (FAPERJ); S\~ao Paulo Research
Foundation (FAPESP) Grants No.~2019/10151-2, No.~2010/07359-6 and
No.~1999/05404-3; Minist\'erio da Ci\^encia, Tecnologia, Inova\c{c}\~oes e
Comunica\c{c}\~oes (MCTIC); Czech Republic -- Grant No.~MSMT CR LTT18004,
LM2015038, LM2018102, LM2023032, CZ.02.1.01/0.0/0.0/16{\textunderscore}013/0001402,
CZ.02.1.01/0.0/0.0/18{\textunderscore}046/0016010 and
CZ.02.1.01/0.0/0.0/17{\textunderscore}049/0008422; France -- Centre de Calcul
IN2P3/CNRS; Centre National de la Recherche Scientifique (CNRS); Conseil
R\'egional Ile-de-France; D\'epartement Physique Nucl\'eaire et Corpusculaire
(PNC-IN2P3/CNRS); D\'epartement Sciences de l'Univers (SDU-INSU/CNRS);
Institut Lagrange de Paris (ILP) Grant No.~LABEX ANR-10-LABX-63 within
the Investissements d'Avenir Programme Grant No.~ANR-11-IDEX-0004-02;
Germany -- Bundesministerium f\"ur Bildung und Forschung (BMBF); Deutsche
Forschungsgemeinschaft (DFG); Finanzministerium Baden-W\"urttemberg;
Helmholtz Alliance for Astroparticle Physics (HAP);
Helmholtz-Gemeinschaft Deutscher Forschungszentren (HGF); Ministerium
f\"ur Kultur und Wissenschaft des Landes Nordrhein-Westfalen; Ministerium
f\"ur Wissenschaft, Forschung und Kunst des Landes Baden-W\"urttemberg;
Italy -- Istituto Nazionale di Fisica Nucleare (INFN); Istituto
Nazionale di Astrofisica (INAF); Ministero dell'Universit\`a e della
Ricerca (MUR); CETEMPS Center of Excellence; Ministero degli Affari
Esteri (MAE), ICSC Centro Nazionale di Ricerca in High Performance
Computing, Big Data and Quantum Computing, funded by European Union
NextGenerationEU, reference code CN{\textunderscore}00000013; M\'exico -- Consejo
Nacional de Ciencia y Tecnolog\'\i{}a (CONACYT) No.~167733; Universidad
Nacional Aut\'onoma de M\'exico (UNAM); PAPIIT DGAPA-UNAM; The Netherlands
-- Ministry of Education, Culture and Science; Netherlands Organisation
for Scientific Research (NWO); Dutch national e-infrastructure with the
support of SURF Cooperative; Poland -- Ministry of Education and
Science, grants No.~DIR/WK/2018/11 and 2022/WK/12; National Science
Centre, grants No.~2016/22/M/ST9/00198, 2016/23/B/ST9/01635,
2020/39/B/ST9/01398, and 2022/45/B/ST9/02163; Portugal -- Portuguese
national funds and FEDER funds within Programa Operacional Factores de
Competitividade through Funda\c{c}\~ao para a Ci\^encia e a Tecnologia
(COMPETE); Romania -- Ministry of Research, Innovation and Digitization,
CNCS-UEFISCDI, contract no.~30N/2023 under Romanian National Core
Program LAPLAS VII, grant no.~PN 23 21 01 02 and project number
PN-III-P1-1.1-TE-2021-0924/TE57/2022, within PNCDI III; Slovenia --
Slovenian Research Agency, grants P1-0031, P1-0385, I0-0033, N1-0111;
Spain -- Ministerio de Econom\'\i{}a, Industria y Competitividad
(FPA2017-85114-P and PID2019-104676GB-C32), Xunta de Galicia (ED431C
2017/07), Junta de Andaluc\'\i{}a (SOMM17/6104/UGR, P18-FR-4314) Feder Funds,
RENATA Red Nacional Tem\'atica de Astropart\'\i{}culas (FPA2015-68783-REDT) and
Mar\'\i{}a de Maeztu Unit of Excellence (MDM-2016-0692); USA -- Department of
Energy, Contracts No.~DE-AC02-07CH11359, No.~DE-FR02-04ER41300,
No.~DE-FG02-99ER41107 and No.~DE-SC0011689; National Science Foundation,
Grant No.~0450696; The Grainger Foundation; Marie Curie-IRSES/EPLANET;
European Particle Physics Latin American Network; and UNESCO.
\end{sloppypar}

%% file: PRL_main.bbl
\begin{thebibliography}{99} 
\def\etal{et~al.\xspace}



\bibitem{ref:UHECRWhitePaper}
A. Coleman \emph{et al.},
\href{https://arxiv.org/abs/2205.05845}{arXiv:2205.05845, (2022).}

\bibitem{ref:augernim}
A. Aab \emph{et al.} (Pierre Auger Collaboration), \href{https://doi.org/10.1016/j.nima.2015.06.058}
     {\emph{Nucl. Instrum. Meth. A} \textbf{798}, 172--213 (2015)}

\bibitem{ref:AERA_Main}
P. Abreu \emph{et al.} (Pierre Auger Collaboration), \href{http://dx.doi.org/10.1088/1748-0221/7/10/P10011}{J. Instrum. \textbf{7}, P10011 (2012).}


\bibitem{ref:histlopes}
H. Falcke \emph{et al.} (LOPES Collaboration), \href{https://doi.org/10.1038/nature03614}
     {\emph{Nature} \textbf{435}, 313--316 (2005).}

\bibitem{ref:MGMR}
O. Scholten, K. Werner, F. Rusydi \href{https://doi.org/10.1016/j.astropartphys.2007.11.012}{Astropart. Phys. \textbf{29}, 94--103 (2008).}

\bibitem{ref:codalema}
D. Ardouin \emph{et al.}, \href{https://doi.org/10.1016/j.astropartphys.2009.01.001}
     {\emph{Astropart. Phys.} \textbf{31}, 192--200 (2009).}

\bibitem{ref:zhaires}
J. Alvarez-Muniz, W. R. Carvalho, Jr., E. Zas, \href{https://doi.org/10.1016/j.astropartphys.2011.10.005}{Astropart. Phys. \textbf{35}, 325--341 (2012).}

\bibitem{ref:MicroCoREAS}
T. Huege, M. Ludwig, C. W. James, \href{https://doi.org/10.1063/1.4807534}{AIP Conf. Proc. \textbf{1535}, 128 (2013).}

\bibitem{ref:2dldf}
A. Nelles \emph{et al.},
\href{https://doi.org/10.1088/1475-7516/2015/05/018}{JCAP \textbf{05}, 018 (2015).}

\bibitem{ref:EnergyScalePRL}
A. Aab \emph{et al.},
\href{https://doi.org/10.1103/PhysRevLett.116.241101}{Phys. Rev. Lett. \textbf{116}, 241101 (2016).}

\bibitem{ref:HuegeReview}
Huege, T., \href{https://doi.org/10.1016/j.physrep.2016.02.001}{Phys. Rept. 620, 1--52 (2016).}

\bibitem{ref:Schroder}
F. G. Schr\"oder, \href{https://doi.org/10.1016/j.ppnp.2016.12.002}{Prog. Part. Nucl. Phys. 93, 1--68 (2017).}

\bibitem{ref:Astro2020}
F. G. Schr\"oder \emph{et al.}, \href{https://doi.org/10.48550/arXiv.1903.07713}{Bull. Am. Astron. Soc. 51, 131 (2019).}

\bibitem{ref:EnergyScalePRD}
A. Aab \emph{et al.},
\href{https://doi.org/10.1103/PhysRevD.93.122005}{Phys. Rev. D \textbf{93}, 122005 (2016).}

\bibitem{ref:hadr_qgs}
S. Ostapchenko,
\href{https://doi.org/10.1016/j.nuclphysbps.2005.07.026}{Nucl. Phys. B Proc. Suppl. \textbf{151}, 143 (2006).}

\bibitem{ref:hadr_epos}
T. Pierog \emph{et al.},
\href{https://doi.org/10.1103/PhysRevC.92.034906}{Phys. Rev. C \textbf{92}, 034906 (2015).}

\bibitem{ref:hadr_sibyll}
F. Riehn \emph{et al.},
\href{https://doi.org/10.1103/PhysRevD.102.063002}{Phys. Rev. D \textbf{102}, 063002 (2020).}

\bibitem{ref:POA_Composition_ICRC2019}
A. Yushkov \emph{et al.} (Pierre Auger Collaboration), 
\href{https://doi.org/10.22323/1.358.0482}{in \emph{Proceedings of the 36th International Cosmic Ray Conference}, \textbf{PoS(ICRC2019)482} (2019).}

\bibitem{ref:SD_energyscale}
B. R. Dawson \emph{et al.} (Pierre Auger Collaboration), \href{https://doi.org/10.22323/1.358.0231}{in \emph{Proceedings of the 36th International Cosmic Ray Conference}, \textbf{PoS(ICRC2019)231} (2019).}

\bibitem{ref:Corsikamain}
D. Heck \emph{et al.}, \href{https://inspirehep.net/literature/469835}{FZKA Tech. Umw. Wis. B \textbf{6019}, (1998).}

\bibitem{ref:gdastool}
P. Mitra \emph{et al.}, \href{https://doi.org/10.1016/j.astropartphys.2020.102470}{Astropart. Phys. \textbf{123}, 102470 (2020).}

\bibitem{ref:gdasinoffline}
P. Abreu \emph{et al.}, \href{https://doi.org/10.1016/j.astropartphys.2011.12.002}{Astropart. Phys. \textbf{35}, 591 (2012).}

\bibitem{ref:magneticmodel}
C. C. Finlay \emph{et al.}, \href{https://doi.org/10.1111/j.1365-246X.2010.04804.x}{Geophysical Journal International \textbf{183}, 1216-1230 (2010).}

\bibitem{ref:LOFARresults1}
S. Buitink \emph{et al.},
\href{https://doi.org/10.1038/nature16976}{Nature \textbf{531}, 70 (2016).}

\bibitem{ref:LOFARresults0}
S. Buitink \emph{et al.},
\href{https://doi.org/10.1103/PhysRevD.90.082003}{Phys. Rev. D \textbf{90}, 082003 (2014).}

\bibitem{ref:AERAXmaxPRD}
A. Abdul Halim \emph{et al.} (Pierre Auger Collaboration), \href{...}{Phys. Rev. D \textbf{[note: the accompanying paper to this PRL][TBD volume]}, [TBD page numbers] (2023).}

\bibitem{ref:Infill_Efficiency_v2}
P. Abreu \emph{et al.} (Pierre Auger Collaboration),
\href{https://doi.org/10.1140/epjc/s10052-021-09700-w}{Eur. Phys. J. C \textbf{81}, 966 (2021).}

\bibitem{ref:Infill_Efficiency_v3}
D. Ravignani \emph{et al.} (Pierre Auger Collaboration), 
in \emph{Proceedings of the 33rd International Cosmic Ray Conference}, 0693 (2013).

\bibitem{ref:FDXmaxSyst}
J. Bellido \emph{et al.} (Pierre Auger Collaboration), 
\href{https://doi.org/10.22323/1.301.0506}{in \emph{Proceedings of the 35th International Cosmic Ray Conference}, \textbf{PoS(ICRC2017)506} (2017).}

\bibitem{ref:CalorimetryHandbook}
C. W. Fabjan, F. Gianotti,
\href{http://dx.doi.org/10.1103/RevModPhys.75.1243}{Rev. Mod. Phys \textbf{75}, 1243 (2003).}


\bibitem{ref:LOFAR2021}
A. Corstanje \emph{et al.}, \href{https://doi.org/10.1103/PhysRevD.103.102006}{Phys. Rev. D \textbf{103}, 102006 (2021).}

\bibitem{ref:SKAxmaxres}
S. Buitink \emph{et al.}, 
\href{https://doi.org/10.22323/1.395.0415}{in \emph{Proceedings of the 37th International Cosmic Ray Conference}, \textbf{PoS(ICRC2021)415} (2021).}


\bibitem{ref:Xmax_TunkaREX}
P. A. Bezyazeekov \emph{et al.} (Tunka-Rex Collaboration), \href{https://doi.org/10.1103/PhysRevD.97.122004}{Phys. Rev. D \textbf{97}, 122004 (2018).}

\bibitem{ref:YakutskXmax2019}
I. Petrov and S. Knurenko,
\href{https://doi.org/10.22323/1.358.0385}{in \emph{Proceedings of the 36th International Cosmic Ray Conference}, \textbf{PoS(ICRC2019)385} (2019).}

\bibitem{ref:LOPESXmax}
F. G. Schr\"oder \emph{et al.} (LOPES Collaboration), 
\href{https://doi.org/10.22323/1.301.0458}{in \emph{Proceedings of the 35th International Cosmic Ray Conference}, \textbf{PoS(ICRC2017)458} (2017).}

\bibitem{ref:AugerTAWGXmax2023}
A. Yushkov \emph{et al.} (Pierre Auger Collaboration and Telescope Array Collaboration), 
\href{https://doi.org/10.22323/1.444.0249}{in \emph{Proceedings of the 38th International Cosmic Ray Conference}, \textbf{PoS(ICRC2023)249} (2023).}


\end{thebibliography}
